\documentclass[pra,twocolumn,aps,superscriptaddress, nofootinbib]{revtex4-1}
\usepackage{amsmath}
\usepackage{amssymb}
\usepackage{graphicx}
\usepackage[usenames,dvipsnames]{color}
\usepackage{braket}
\usepackage{bm}
\usepackage{subfigure}
\usepackage{times}
\usepackage{hyperref}
\usepackage{pdfpages}
\usepackage{float}
\usepackage{lipsum}
\usepackage[dvipsnames]{xcolor}
\usepackage{comment}
\makeatletter
 \AtBeginDocument{\let\LS@rot\@undefined}
\makeatother

\begin{document}
\title{Dipolar mixtures in checker-board optical bilayers}

\author{Rukmani Bai}
\email{rukmani.bai@itp.uni-hannover.de}
\affiliation{Institut f\"ur Theoretische Physik, Leibniz Universit\"at Hannover,
Appelstrasse 2, D-30167 Hannover, Germany}

\author{Luis Santos}
\email{santos@itp.uni-hannover.de}
\affiliation{Institut f\"ur Theoretische Physik, Leibniz Universit\"at Hannover,
Appelstrasse 2, D-30167 Hannover, Germany}

\begin{abstract}
Ultra-cold dipolar mixtures in component-dependent optical potentials constitute an interesting platform for the study of the interplay between intra- and inter-component anisotropic long-range  interactions. We study the particular case of binary dipolar mixtures placed in separated bilayers which are displaced in an anti-magic wavelength configuration. Using a combination of second-order perturbation theory and cluster-Gutzwiller calculations, we unveil a rich landscape of possible crystalline phases for the two components, showing that, interestingly, inter-site hopping may result, via super-exchange, in solid-into-solid transitions between different crystalline phases.  These crystalline phases and the corresponding transitions can be experimentally realized using e.g. lanthanide mixtures in optical lattices.
\end{abstract}

\maketitle

\section{Introduction} 
\label{sec:Introduction}

Recent years have shown a surge in the interest on dipolar quantum gases, formed by particles with large magnetic or electric dipole moments. Due to the anisotropic long-range character of the dipole-dipole interaction, these gases present a qualitatively new physics compared to the non-dipolar counterparts~\cite{Lahaye2009,Baranov2012,Chomaz2022}. Milestone experiments on dipolar Bose-Einstein condensates have revealed a very rich physics, including roton-like excitations~\cite{chomaz2018}, dipolar droplets~\cite{Kadau2016, Zhang2026}, and dipolar supersolidity~\cite{Tanzi2019, Boettcher2019, Chomaz2019}. 

Dipolar gases in optical lattices provide an excellent platform for the study of the interplay of short- and long-range interactions in many-body lattice models. In particular, itinerant dipoles in optical lattices or tweezer arrays allow for the quantum simulation of extended Hubbard models~\cite{Lahaye2009,Baranov2012, baier_16}, characterized by strong inter-site interactions which may extend well beyond nearest-neighboring sites. These interactions are expected to result in a rich ground-state physics and non-equilibrium dynamics. Recent experiments using highly magnetic erbium atoms have recently observed for the first time the formation of density waves with checker-board or stripe configurations~\cite{Su2023}, which stem directly from the anisotropic inter-site dipole-dipole interaction.


Recent experiments on dipolar mixtures open intriguing possibilities for the study of the competition between intra- and inter-component dipolar interactions~\cite{dePaz2013,trautmann_18, Durastante2020, Alaoui2022, Politi2022, Claude2024, Barral2024, Lecomte2025, Laupetre2026, nie2026}. Interestingly, experiments on lanthanides allow for component-selective optical addressing~\cite{Claude2024}, allowing for the study of scenarios in which the two components experience a different optical lattice. Particularly interesting is the case of binary mixtures in a one-dimensional anti-magic wavelength configuration, in which one component experiences a one-dimensional lattice displaced by half a unit cell compared to the other component, which may be employed e.g. for the realization of frustrated lattice models~\cite{Miotto2026}. 

\begin{figure}[t]
    \centering
    \includegraphics[width=1.0\linewidth]{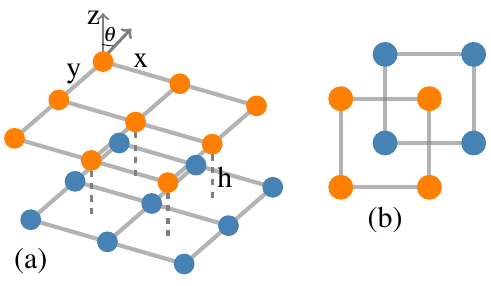} 
    \caption{(a) Scheme of the considered bilayer square-lattice setup. The layers are separated by a vertical distance ${h}$. Each layer is occupied by only one of the components. (b) Scheme of the considered $2 \times 2$ cluster, with $4$ sites in each layer, employed in our CGW calculations.}
    \label{fig:schm}
\end{figure}

In this paper, we consider the particular case of a 2D anti-magic wavelength~(checker-board) configuration, in which each component is placed in a square lattice, but one lattice is displaced half unit cell in each direction on the lattice plane.
Moreover, the two lattices have a vertical displacement, forming a bilayer system, see Fig.~\ref{fig:schm}~(a). By a combination of second-order perturbation theory and cluster Gutzwiller ansatz calculations, we show that this particularly simple arrangement reveals 
for hard-core dipolar bosons
a rich ground-state landscape characterized by different crystalline geometries that depend on the dipole orientation and the inter-layer separation. Interestingly, the tunneling itself may induce a solid-into-solid transition between two different crystalline phases, as a result of super-exchange processes.

The paper is organized as follows. In Sec. \ref{sec:Model}, we introduce the model 
and the cluster Gutzwiller method. Section \ref{sec:Perpendicular} discusses the phase diagram when the dipoles are oriented perpendicular to the layer planes. In Sec.~\ref{sec:Angle} we discuss the case in which the dipoles have a component on the layer planes. Finally, in Sec.~\ref{sec:Conclusions}, we summarize our conclusions.



\begin{figure*}[t]
    \includegraphics[width=0.4\linewidth]{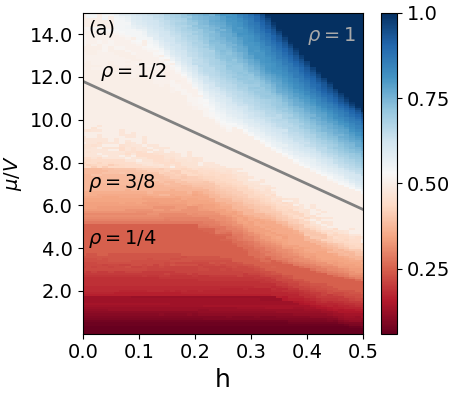}
    \includegraphics[width=0.53\linewidth]{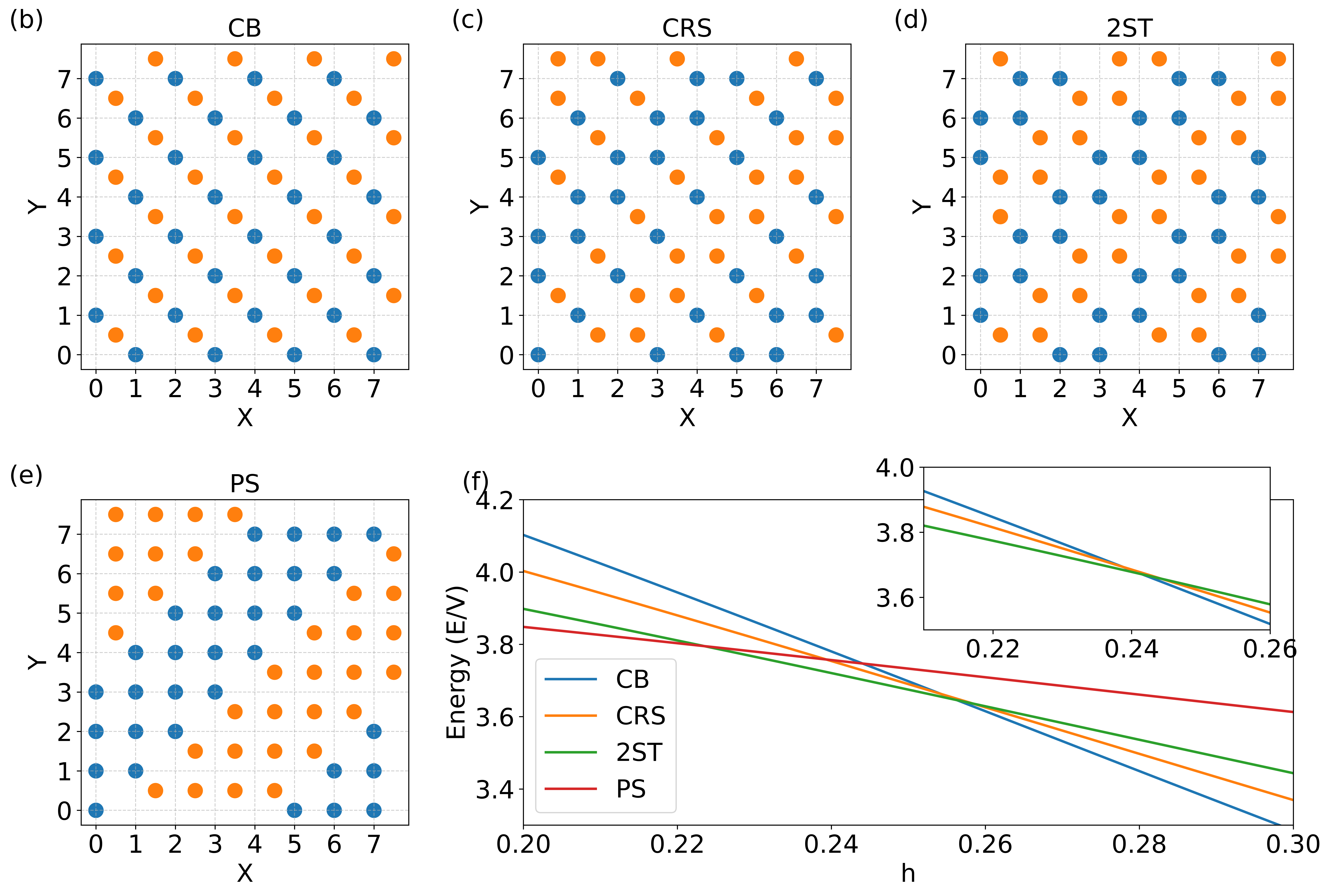}
     \caption{(a) Density $\rho_{1,2}=\rho$ as a function of the interlayer spacing $h$, and the chemical potential $\mu$. We focus on $\rho = 1/2$, considering chemical potentials along the solid line. (b) Checker-board~(CB) crystal. (c) Intermediate crystalline~(CRS) phase. (d) Two-row stripe~(2ST) phase; (e) Phase separated~(PS) phase: here it is limited due to the $8\times 8$ lattice with periodic boundary conditions; larger lattices show the formation of broader domains. (f) Energy of the different crystalline phases in the atomic limit~($J=0$) as a function of $h$. In the inset, we show the energy of the CB, CRS and 2ST phases as a function of $h$ for $J/V=0.2$, showing that, also at finite hopping, the three phases remain approximately degenerate at the CB-2ST transition.}
    \label{fig:ph_dig}
\end{figure*}


\begin{figure}[t]
   \includegraphics[width=1.0\linewidth]{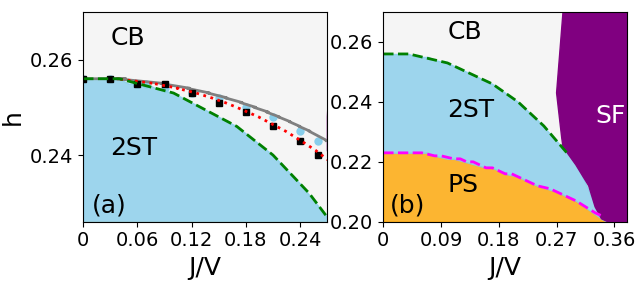}
    \includegraphics[width=0.80\linewidth]{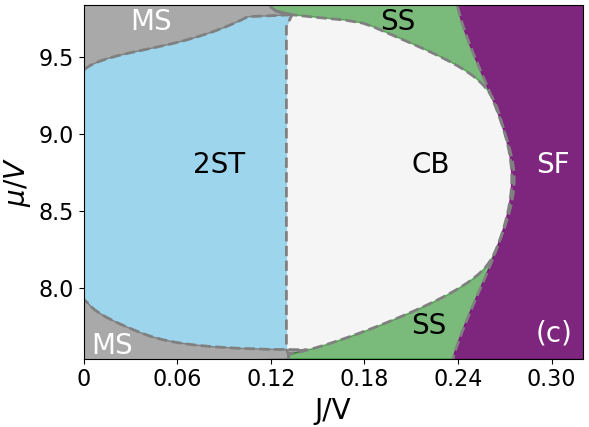}
\caption{(a) CB-2ST transition. The dashed line indicates the CB-2ST transition as obtained from the second-order perturbation theory calculation. As expected, see text, the second-order results with $J_{\mathrm{eff}}=2J$~(solid curve) match with the CGW calculation for a $2\times 2$ lattice~(filled circles), whereas the CGW results for a $4\times 2$ lattice~(filled squares) match with the second-order results using $J_{\mathrm{eff}}=1.6J$~(dotted curve). This result shows both that the second-order perturbation theory is valid at the $J$ values considered, and that CGW cannot be employed to determine the solid-into-solid transitions. (b) Phase diagram as a function of the hopping $J$ and the inter-layer separation $h$, assuming the dipoles oriented perpendicularly to the layer ~($\theta=0$). (c) Phase diagram for $h=0.25$ as a function of $J$ and the chemical potential $\mu$; we depict the lobe of $\rho=1/2$ and its vicinity. The lobe presents a super-exchange-induced solid-into-solid 2ST-CB transition. Associated to this transition there is a change in the nature of the phases outside the lobe, which include a supersolid~(SS) checker-board phase. }
    \label{fig:hopping}
\end{figure}


\section{Model and Method}
\label{sec:Model}

We consider a binary mixture of hard-core dipolar bosons in a component-dependent checker-board-like bilayer set up as that of Fig.~\ref{fig:schm}~(a). 
Component $\kappa={1,2}$ remains confined in layer $\kappa$. The layers are separated by a distance $h$ along the $z$ axis. Both components experience a square optical lattice~(we consider the lattice constant as the length unit), but 
they are in anti-magic wavelength configuration such that 
the lattice of layer $2$ is shifted by half a lattice constant in each direction compared to that of layer $1$, see Fig.~\ref{fig:schm}~(a). The system is described by the extended Bose-Hubbard Hamiltonian:
\begin{eqnarray}
\label{hamil_terms}
\hat{H}& = &-J\sum_{\langle i, j\rangle ,\kappa} \hat{b}_{i,\kappa}^\dagger
                      \hat{b}_{j,\kappa} - \sum_{i, \kappa} 
		      \mu_\kappa \hat{n}_{i,\kappa} \nonumber\\
&& + \frac{1}{2}\sum_{i \ne j, \kappa}   
	      V_{\kappa,\kappa}^{i,j} \hat{n}_{i,\kappa} \hat{n}_{j,\kappa}
        + \sum_{i, j} V_{1,2}^{i,j} \hat{n}_{i,1} \hat{n}_{j,2}, 
\end{eqnarray}
where $\kappa = \{1,2\}$, $\hat b_{j,\kappa}^\dag$~($\hat b_{j,\kappa}$) is the creation~(annihilation) of a boson at site $j$ of layer $\kappa$,   
$\hat{n}_{j,\kappa}=
\hat b_{j,\kappa}^\dag
\hat b_{j,\kappa}$, $(\hat b_{j,\kappa}^\dag)^2=0$, 
$J$ is the nearest-neighbor intra-layer hopping rate,
which we assume for simplicity to be the same for both components, and $\mu_{\kappa}$ is the chemical potential for component $\kappa$. Assuming that an external field orients the dipoles on the $xz$ plane 
along the direction ${\mathbf{e}_d}=(\sin\theta,0,\cos\theta)$, 
the dipole-dipole interaction between site $i$ at layer $\kappa$ and site $j$ at layer $\kappa'$ 
is given by 
\begin{eqnarray}
V_{\kappa,\kappa'}^{i,j} \!=\! \frac{V_{\kappa,\kappa'}}{|{\mathbf r}_{i,\kappa}- \mathbf{r}_{j,\kappa'}|^3} 
\!
\left ( 1 - 3 
\frac{\left (\mathbf{e}_d\cdot \left ({\mathbf r}_{i,\kappa}- \mathbf{r}_{j,\kappa'}\right ) \right )^2}{|{\mathbf r}_{i,\kappa}- \mathbf{r}_{j,\kappa'}|^2}
\right ).
\end{eqnarray}
We assume for simplicity $V_{1,1}=V_{2,2}=V_{1,2}=V$, which we consider our energy scale, and assume $\mu_1=\mu_2=\mu$ such that both components have equal population. 
Here ${\mathbf r}_{j,1} = (j_x, j_y, 0)$ and ${\mathbf r}_{j,2} = (j_x+1/2, j_y+1/2, h)$.

We are interested in the ground-state properties of the two-component bilayer system. The transition between superfluid and insulating phases may be well studied by means of cluster Gutzwiller~(CGW) ansatz calculations~\cite{bai_18,bandyopadhyay_22,luhmann_13}~(tunneling-induced solid-into-solid transitions are in contrast not well resolved using CGW, as
discussed below). 
In CGW, the square lattice 
is split in each layer into clusters of $M\times N$ sites~(along $x$ and $y$, respectively). 
The ground-state is written as product state of clusters: $|\Psi_{\mathrm{CGW}}\rangle=\bigotimes_c |\psi\rangle_{c} $, where the cluster state is expressed as a linear combination of Fock states: 
$|\psi\rangle_c = \sum_{\mathbf{n},\mathbf{m}}f^{c}_{\mathbf{n},\mathbf{m}}|\mathbf{n},\mathbf{m}\rangle_{c}$, 
where $\mathbf{n} =\{n_1, n_2, \ldots, n_{M\times N}\}$ for the first species and  $\mathbf{m} =\{m_1, m_2, \ldots, m_{M\times N}\}$ for the second one. 
We then approximate: $\hat H \simeq \sum_c \left (\hat H_c + \hat H_c^{\mathrm{MF}}\right )$, such that 
\begin{eqnarray}
\hat{H_c}& = &-J\sum_{\langle i, j\rangle \in c,\, \kappa} 
\!\!\!\!
\hat{b}_{i,\kappa}^\dagger
                      \hat{b}_{j,\kappa} - \sum_{i\in c, \kappa} 
		      \mu_\kappa \hat{n}_{i,\kappa} \nonumber\\
&& + \frac{1}{2}\sum_{(i \ne j) \in c, \,\kappa} \!\!\!\!  
	      V_{\kappa,\kappa}^{i,j} \hat{n}_{i,\kappa} \hat{n}_{j,\kappa}
        + \sum_{(i, j)\in c} \!\! V_{12}^{i,j} \hat{n}_{i,1} \hat{n}_{j,2}, 
\end{eqnarray}
is the exact Hamiltonian within the cluster $c$~\cite{footnote-1}, 
whereas the hopping and interactions with neighboring clusters are considered in mean-field:
\begin{eqnarray}
\hat{H}_c^{\mathrm{MF}}& = &\sum_{i\in c, \, \kappa} \left [-J\eta_{i,\kappa} (\hat{b}_{i,\kappa}^\dagger
+ \hat{b}_{i,\kappa}) 
+ \beta_{i,\kappa} \hat n_{i,\kappa}
\right ],
\label{eq:HcMF}
\end{eqnarray}
where $\eta_{i,\kappa}\equiv \sum_{\langle j\in c'\neq c\rangle_i}\phi_{j,\kappa}$~(with $\langle j\in c'\neq c\rangle_i$ denoting the neighboring sites to site $i$ which do not belong to cluster $c$), 
$\beta_{i,\kappa}\equiv\sum_{j\in c'\neq c}\left ( V_{\kappa,\kappa}^{i,j}\rho_{j,\kappa} + V_{1,2}^{i,j}\rho_{j,\kappa'\neq \kappa}\right )$, and we have introduced the superfluid order parameter $\phi_{j,1}= \langle\Psi_{\rm CGW}|\hat{b}_{j,1}|\Psi_{\rm CGW}\rangle 
= \sum_{\mathbf{n},\mathbf{m}}\sqrt{n_j} f^{c\, *}_{\mathbf{n}-\mathbf{e}_j, \mathbf{m}} f^{c}_{\mathbf{n},\mathbf{m}}$, 
and the average occupancy 
$\rho_{j,1} = \langle\Psi_{\rm CGW}|\hat{n}_{j,1}|\Psi_{\rm CGW}\rangle
= \sum_{\mathbf{n},\mathbf{m}} n_j |f^{c}_{\mathbf{n},\mathbf{m}}|^2$, and similarly for the component $2$.
We perform our CGW calculations on a bilayer system with $8 \times 8$ lattice sites per layer, which we split into $2 \times 2$ clusters, i.e. each cluster has $8$ sites, $4$ in each layer, see Fig.~\ref{fig:schm}~(b). We consider periodic boundary conditions along both $x$ and $y$. 


\section{Dipoles perpendicular to the layers}
\label{sec:Perpendicular}

We start our analysis with the case in which the dipoles are aligned along the $z$-direction. We focus first on the atomic limit, $J = 0$, and incorporate below the non-trivial effect of hopping. Figure~\ref{fig:ph_dig}~(a) depicts the lattice filling on each layer ($\rho_{1,2}=\rho$) as a function of the chemical potential $\mu$ and 
the interlayer separation $h$. Increasing $\mu/V$ leads to an $h$-dependent sequence of different lattice fillings. In this paper, we focus on half filling, $\rho = 1/2$, and hence for each value of $h$ we consider a chemical potential along the solid line in Fig.~\ref{fig:ph_dig}~ (a). 

In the atomic limit, the ground-state is a Fock state with given occupations $n_{i,\kappa}$, which minimize the interaction energy:
\begin{equation}
E = \frac{1}{2}\sum_{i \ne j, \kappa} V_{\kappa,\kappa}^{i,j}n_{i,\kappa} n_{j,\kappa}+ \sum_{i, j} V_{1,2}^{i,j} n_{i,1} n_{j,2}.
\label{energy}        
\end{equation}
The competition between intra- and inter-species interactions leads to the emergence of various crystalline phases, with characteristic occupation patterns and vanishing superfluid order parameter.  For sufficiently weak inter-species interaction, i.e. sufficiently large separation $h$, the ground-state is characterized by a checker-board-like filling of each layer.
In this checker-board~(CB) phase, the population of both components align, forming a characteristic filament-like pattern, see Fig.~\ref{fig:ph_dig}~(b). 
As $h$ decreases the inter-species interaction strength increases, and the ground-state of the binary mixture undergoes a transition to a crystalline phase characterized by alternated two-rows-wide stripes of each component. This two-row stripe~(2ST) phase is sketched in Fig.~\ref{fig:ph_dig}~(d). 
Interestingly, as shown in Fig.~\ref{fig:ph_dig}~(c), at the transition, the CB and 2ST phases are quasi-degenerate~(the energy difference is $\Delta E\simeq 0.004V$), with yet a third crystalline arrangement, which we denote as the intermediate crystalline~(CRS) phase. This crystal presents the intricate geometry depicted in Fig.~\ref{fig:ph_dig}~(c). 
Finally, for sufficiently small separation $h$, the repulsive inter-species interaction becomes significantly stronger than the intra-species interaction, and the system transitions into a phase-separated~(PS) state, see Fig. \ref{fig:ph_dig}~(e). 
We should point that the 
results were evaluated for a relatively small $8\times 8$ lattice. For large system sizes, 
the 2ST-PS transition shifts and occurs via other intermediate stripe phases (e.g. for a 
$16 \times 16$ lattice, we observe a very narrow intermediate region~(in between 2ST and PS) with a stripe phase formed by 
alternated four-row stripes of each component).

A finite particle hopping eventually results not only in crystal melting into a superfluid phase, but also in solid-into-solid transitions (i.e. transitions between different crystal geometries), induced by super-exchange processes~\cite{anderson_50, mendes_23}. The latter result from virtual hops, which affect in a different way the various crystal geometries. For each one of the phases 
depicted in Figs.~\ref{fig:ph_dig}(b--e), we evaluate the energy shift $\Delta E$ associated with the hopping of a single particle into an empty neighboring site. This virtual hopping process results in a second-order energy shift $-J^2/\Delta E$. Adding up the contribution of all particles, we evaluate the overall change in the energy per particle for the phase $\alpha=$ CB, ST, CRS, and PS, from its atomic limit value $E_{\alpha;0}(h, V)$ to $E_{\alpha}(h, V, J) = E_{\alpha,0}(h, V) - \gamma_\alpha(h,V) J^2/V$
, for details see Appendix~\ref{spt}.  Super-exchange induces the shift of the phases, resulting in tunneling-induced CB-2ST and 
PS-2ST transitions, as depicted in Fig.~\ref{fig:hopping}. 

Solid-into-solid transitions are not well reproduced by 
CGW methods. This is because, hops between different clusters 
are evaluated in mean-field, and are hence canceled in crystalline phases~($\eta_{i,\kappa}=0$ in Eq.~\eqref{eq:HcMF}). As a result, calculations with $2\times 2$~($4\times 2$) clusters, are characterized by an effective coordination number $z_{\mathrm{eff}}=2$~($2.5$), and not of $z=4$. This point is illustrated in Fig.~\ref{fig:hopping}~(a), where we show that the evaluated 2ST-CB boundary for $2\times 2$~($4\times 2$) clusters coincides with the correct results evaluated in second-order perturbation, when employing $J_{\mathrm{eff}}=2$~($1.6$)~$J$, see Appendix~\ref{effective}. 

Very large clusters would be necessary to reproduce the actual solid-to-solid transition, rendering the CGW treatment of the tunneling-induced solid-into-solid transitions 
numerically unfeasible. We hence evaluate those transitions using the super-exchange calculation. In contrast  
this problem does not affect significantly the crystal-to-superfluid transitions, and hence we use CGW to determine the melting transitions
into a superfluid phase, characterized by homogeneous density and finite superfluid order parameter. Using this combination of methods we obtain the phase diagram depicted in Fig.~\ref{fig:hopping}~(b). We employ this procedure in all other results discussed below.

Interestingly, although the intermediate CRS phase is strictly speaking never the ground state, its energy remains at finite hopping quasi-degenerate with that of the CB and 2ST crystals in the vicinity of the CB-2ST transition~(see inset of Fig.~\ref{fig:ph_dig}~(f)). Indeed, we observe this phase in our CGW calculation, and we expect that it would appear as well as metastable solution in experiments.

The hopping-induced solid-to-solid transition results in a distortion of the insulator $\rho=1/2$ lobe in the $(J,\mu)$ phase diagram, as depicted in Fig.~\ref{fig:hopping}~(c). Upon leaving the 
$\rho=1/2$ lobe, the 2ST phase  transitions into a regime of metastable~(MS) states , characterized by no superfluid parameter or density order, and by a fluctuating density pattern from run to run~(we perform simulations with up to $10$ different random initial conditions). The CB phase transitions into a SF or a checker-board supersolid~(SS) phase. The latter phase is characterized by a modulated density and finite superfluid order parameter.


\section{Anisotropic dipolar interaction with polarization angle}
\label{sec:Angle}

\begin{figure}[t!]
    \includegraphics[width=0.95\linewidth]{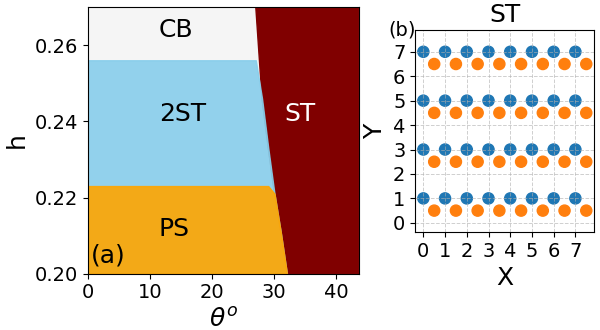}
     \caption{(a) Phase diagram in the atomic limit ($J=0$) as a function of the dipole orientation $\theta$ and the inter-layer spacing $h$. (b) Stripe~(ST) phase characteristic of large-enough angles $\theta$.}
     \label{fig:angle}
\end{figure}

Up to this point we have only considered dipoles along $z$~($\theta=0$). For $\theta\neq 0$, the dipoles have a projection along $x$, hence breaking the isotropy of the $xy$ plane. Figure \ref{fig:angle}~(a) shows the phase diagram as a function of the separation $h$ and the orientation angle $\theta$ in the atomic limit, $J/V \to 0$. 
When $\theta$ grows the intra-component interaction 
along $x$ becomes eventually attractive for $\theta>\arcsin (1/\sqrt{3})$, while remaining repulsive along $y$. As a result, the 
ground-state undergoes a phase transition at a critical $\theta_{\mathrm{c}}(h)$ from the isotropic ground-state crystals discussed in the previous section~(PS, 2ST, CB) into the stripe~(ST) phase, where particles align in stripes along $x$, formed by two rows, one of each component, see Fig.~\ref{fig:angle}~(b). 

\begin{figure}[t]
    \includegraphics[width=0.6\linewidth]{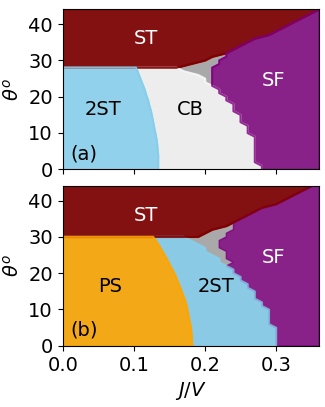}
    \caption{Phase diagram as a function of the hopping $J$ and the orientation angle $\theta$ for $h = 0.25$~(a) and $h = 0.217$~(b). Note the solid-into-solid transitions, as well as the significant dependence of the crystal-to-superfluid transition depending on the orientation angle. The gray zone indicates amorphous metastable solutions in our CGW calculations.}
    \label{fig:angle_j}
\end{figure}

Hopping results, also for $\theta\neq 0$, in solid-into-solid 2ST-CB and PS-2ST transitions for $\theta<\theta_c(h)$, see
Figs~\ref{fig:angle_j}, which we obtain again employing second-order perturbation theory.
Due to the hopping induced transition, at $\theta_{\mathrm{c}}$ the system presents a triple point ST-2ST-CB~(ST-PS-2ST) point. Increasing $J/V$ eventually drives a $\theta$-dependent crystal melting into a superfluid. The crystal-to-superfluid transition presents a marked $\theta$-dependence. At the critical $\theta_{\mathrm{c}}(h)$ the SF penetrates down to small $J/V$ values. In the triple point region CB-ST-SF~(2ST-ST-SF) our CGW results show the appearance of  metastable states, with random density distribution and finite superfluid order parameter (grey areas in the figures). 


%




\section{Conclusion}
\label{sec:Conclusions}

Ultracold dipolar mixtures in component-dependent optical potentials open interesting perspectives for the study of the rich physics that results from the 
interplay between intra- and inter-component interactions. Using a combination of second-order perturbation theory and cluster-Gutzwiller calculations, we have analyzed the case of binary mixtures in a bilayer set-up with optical lattices in anti-magic wavelength~(checker-board) configuration. Whereas, as recently observed experimentally~\cite{Su2023}, single-component dipoles in square lattices present a transition from checker-board to stripe phases as a function of the dipole orientation, the ground-states of the considered bilayer geometry depend both on the dipole orientation and on the separation between the layers. This results in a richer landscape of possible crystalline phases for the two components, see Figs.~2 (b-e) and Fig. 4(b). Interestingly, the hopping itself may result, via super-exchange, in solid-into-solid transitions between different crystalline phases.  These crystalline phases and the corresponding transitions can be experimentally realized using ultra-cold lanthanides in optical lattices~\cite{baier_16,Su2023}.Lanthanide mixtures are not only dipolar, but also allow for the creation of component-selective optical potentials, due to their available narrow optical transitions~\cite{Claude2024}. A Dy/Dy mixture, employed in very recent experiments~\cite{Durbeck_26, nie2026} would allow for $V_{11}=V_{22}=V_{12}$, although in our work, we have chosen this for simplicity.

Finally, we point that the anti-magic-wavelength~(checker-board) bilayer potentials discussed in this paper only illustrate partially the possibilities of a dipolar bilayer set up. Particularly interesting would be to extend ideas recently developed concerning the generation of twisted-bilayer geometries~\cite{Gonzalez2019,Meng2023} to the dipolar case, which may permit the study of the interplay between twisting and dipolar interactions.

\acknowledgements
We acknowledge the support of the Deutsche Forschungsgemeinschaft (DFG, German Research Foundation) -- Project-ID 274200144 -- SFB 1227 DQ-mat within the project A04, and under Germany's Excellence Strategy -- EXC-2123 Quantum-Frontiers -- 390837967.

\appendix

\section{Second Order energy correction}
\label{spt}
In this Appendix, we describe the details of the second order perturbation theory used in the paper.

In the atomic limit $J=0$, the ground
state is a Fock state $|\Psi\rangle = |\{n_{i,\kappa}\}\rangle$, which corresponds to 
one of the crystalline phases discussed in the text, 
with an interaction energy
\begin{equation}
\begin{aligned}
E_0(\{n_{i,\kappa}\})
&=
\frac{1}{2}
\sum_{\kappa=1,2}
\sum_{i,j\neq i}
V_{\kappa,\kappa}^{i,j}
n_{i,\kappa}n_{j,\kappa} \\
&\quad+
\sum_{i,j}
V_{1,2}^{i,j}
n_{i,1}n_{j,2}.
\end{aligned}
\end{equation}
In order to evaluate the second order energy correction we first operate with the hopping Hamiltonian on the crystalline state:
\begin{equation}
\hat{H}_T|\Psi\rangle
=
-J
\sum_{\kappa}
\sum_{i,\delta}
|\Psi_{\kappa,i,\delta}\rangle,
\end{equation}
where $|\Psi_{\kappa,i,\delta}\rangle$ is the state in which a single particle in layer $\kappa$ has hopped from site $i$ to site $i+\delta$. The sum goes
over all sites which are occupied in the ground state.

In order to evaluate the energy difference associated with a single hop, we first evaluate for the original crystalline structure the interaction energy of the particle at site $i$ in layer $\kappa$ with all other particles:
\begin{equation}
E'_{i,\kappa} = \sum_{j\neq i}
V\left({\bf r}_{j,\kappa}-{\bf r}_{i,\kappa}\right)
+ \sum_j V\left({\bf r}_{j,\kappa\,'\neq \kappa}-{\bf r}_{i,\kappa}
\right),
\end{equation}
and then calculate the interaction energy of the particle when it has hopped into a neighboring 
empty place ${\bf r}_{i+\delta,\kappa}$:
\begin{align}
E''_{i+\delta,\kappa}
&=
\sum_j V\!\left(\mathbf{r}_{j,\kappa}-\mathbf{r}_{i+\delta,\kappa}\right)\nonumber \\
&\quad + \sum_j V\!\left(\mathbf{r}_{j,\kappa\,'\neq \kappa}-\mathbf{r}_{i+\delta,\kappa} \right)\nonumber\\
&\quad - V\left(\mathbf{r}_{i+\delta,\kappa}-\mathbf{r}_{i,\kappa}\right).
\end{align}
where again the sums go over the sites occupied in the original crystalline state. The second-order energy contribution of that hop is then:
$\Delta E_{i,\delta,\kappa} =-J^2/dE_{i,\delta,\kappa}$, with $ dE_{i,\delta,\kappa} = E{''}_{i+\delta,\kappa}-
E'_{i,\kappa}$. The total energy correction of the crystalline state is hence given by adding the contribution of all possible hops:
\begin{equation}
\Delta E = \sum_{\kappa}\sum_{i,\delta} \Delta E_{i,\delta,\kappa}.
\end{equation}
Hence, the energy of the crystalline state in the presence of weak hopping may be approximated as
$E = E_0 + \Delta E$.
We consider that the crystalline structure is such that $E'_{i,\kappa} = E'$ is the same for all sites~(this is so for CB, 2ST and PS phases). We can then write
\begin{equation}
E_0 = \frac{1}{2}\sum_{\kappa}\sum_i E'_{i,\kappa}
= N E',
\end{equation}
with $N$ the number of particles in one layer, which we consider to be the same in both layers.
Further, the energy change due to any hop is for CB, 2ST and PS also the same $dE_{i,\delta,\kappa} = dE$.
Then the energy correction due to all hops is
\begin{equation}
\Delta E = -2NZ\frac{J^2}{dE},
\end{equation}
with $Z$ the number of empty neighboring sites in the same layer in the original crystalline state.
We may now evaluate the total energy per particle of the crystalline phase:
%
\begin{equation}
\epsilon = \frac{E}{2N} = \frac{E'}{2} - Z\frac{J^2}{dE}
\end{equation}
Different crystalline phases present different $E'$, $dE$, and $Z$. In our calculation we calculate $E'$ and $dE$ numerically for a fixed interlayer separation $h$.
For example for $h = 0.25$
%
\begin{eqnarray}
\epsilon_{CB} &\simeq & 
3.70 - 1.22\frac{J^2}{V}, \\
\epsilon_{2ST} &\simeq& 3.67 - 0.58\frac{J^2}{V}.
\end{eqnarray}
The CB-2ST transition is given by $\epsilon_{\mathrm{CB}} = \epsilon_{\mathrm{2ST}}$, which leads, when evaluated for different $h$ values, to the figure depicted in Fig.~\ref{fig:hopping}.
The transition between 2ST and PS phase can be evaluated in the same way. A similar procedure can be applied to the CRS phase as well, although in that case the procedure is more involved, since each particle and particle hop must be considered separately, since not all of them are characterized by the same energy.


\section{Effective coordination number}
\label{effective}

\begin{figure}[t]
    \includegraphics[width=0.95\linewidth]{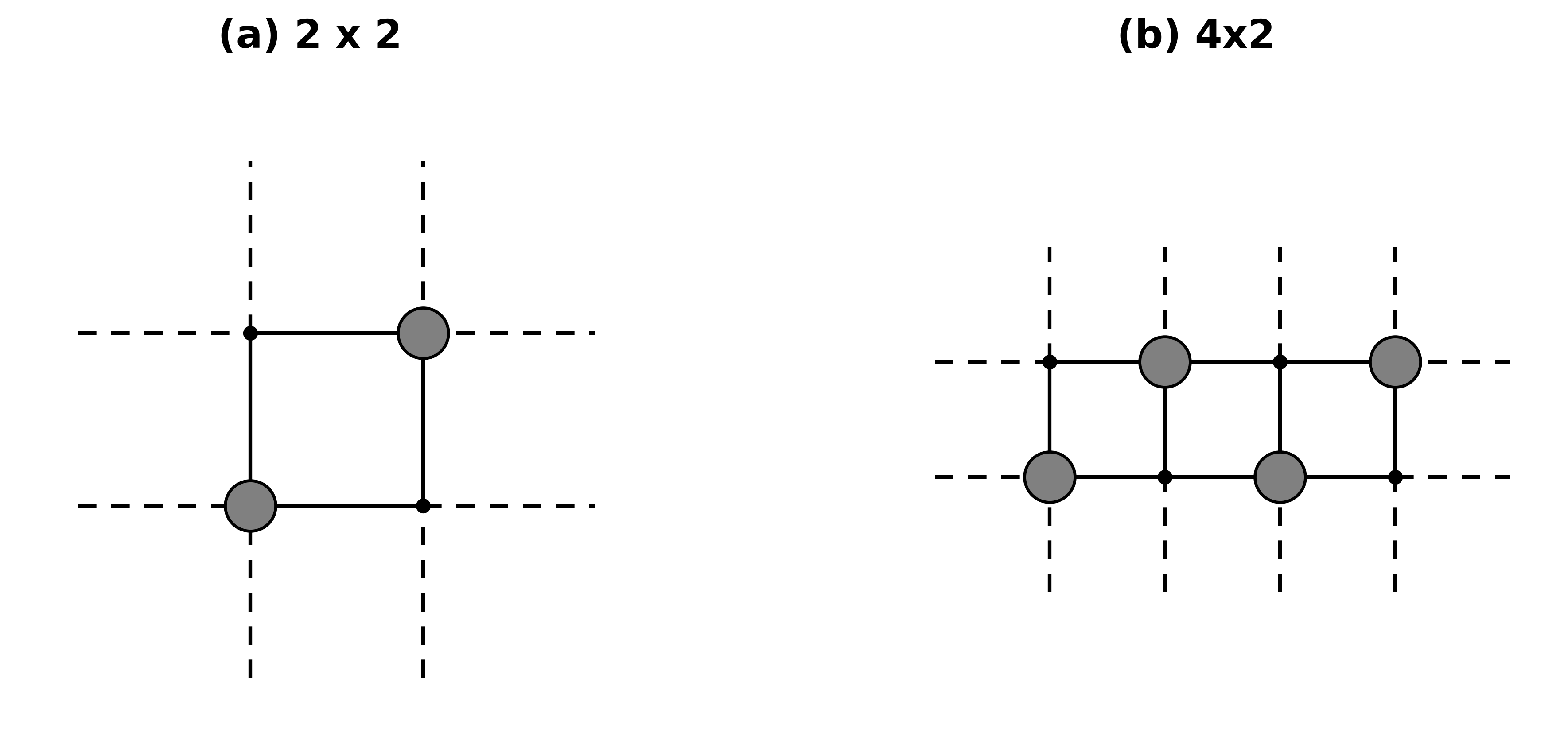}
     \caption{Suppression of hopping for CB phase for (a) $2 \times 2$ cluster (b) $ 4 \times 2$ cluster}
    \label{fig:cluster1}
\end{figure}

The reasoning behind the effective coordination number introduced in Sec.~\ref{sec:Perpendicular} may be well understood for the CB case from Fig.~\ref{fig:cluster1}. 
 In Fig.~\ref{fig:cluster1}(a) we depict the case of a $2\times 2$ cluster. Note that the particles may hop to the four neighbors ($z=4$). However, in the cluster Gutzwiller approach, hops out of the cluster~(dashed lines) are accounted in mean-field theory. For crystalline phases, the corresponding mean-field contribution vanishes. As result, the hops along the dashed lines vanish in this formalism. The effective coordination number is hence reduced to $z_{\rm eff}=2$. A similar reasoning may be applied to Fig.~\ref{fig:cluster1}(b), where we consider the case of a $4\times 2$ cluster. Removing the hops along the dashed lines, result in an average effective coordination number $z_{\rm eff}=2.5$. Applying the same reasoning to the 2ST crystal results in the same effective coordination numbers.

 Due to this effective coordination number, if we want to retrieve the same hopping per particle, the hopping must be changed into $J_{\rm eff}$, such that $z_{\mathrm{eff}}J_{\mathrm{eff}} = zJ$, with $z=4$. Thus, for the $2\times2$~($4\times2$) clusters, we obtain $J_{\mathrm{eff}} = 2(1.6)J$, as discussed in Sec.~\ref{sec:Perpendicular}.

\newpage

\bibliography{ref}{}

\begin{thebibliography}{31}
\bibitem [{foo()}]{footnote-1}%
  \BibitemOpen
  \href@noop {} {}\bibinfo {note} {In our CGW calculations, we determine the
  dipole-dipole interaction between two sites as given by the shortest
  distance, taking into account periodic boundary conditions}\BibitemShut
  {NoStop}%
\end{thebibliography}

\end{document}